\magnification=1200
\baselineskip=22pt plus 1pt minus 1pt
\tolerance=1000
\parskip=0pt
\parindent=25pt



\font\tenss=cmss10


\font\tenrm=cmr10

\font\eightit=cmti8


\def\hspace{~\hskip 1cm ~}

\def\next{~~~,~~~}

\def\N{\kappa}

\def\m{ {\vec m} }
\def\n{ {\vec n} }
\def\vk{ {\vec k} }
\def\vl{ {\vec l} }

\def\vg{ {\vec g} }
\def\vh{ {\vec  h} }

\def\vtheta{ {\vec \theta} }

\def\vgamma{ {\vec  \gamma} }
\def\vone{ {\vec  1} }

\def\bU{ U \kern-.75em\raise.65em\hbox{$-$} }
\def\bW{ W \kern-.9em\raise.65em\hbox{$-$} }
\def\bUq{ U \kern-.75em\raise.65em\hbox{$-${\eightit q}} }
\def\bUqj{ U_j \kern-1.05em\raise.65em\hbox{$-\,${\eightit q}} }
\def\bUqone{ U_1 \kern-1.05em\raise.65em\hbox{$-\,${\eightit q}} }
\def\bUqtwo{ U_2 \kern-1.05em\raise.65em\hbox{$-\,${\eightit q}} }
\def\bJ{{\bar J}}
\def\bq{{ \bar q}}

\def\N{ \kappa }

\def\zz{ \hbox{\tenss Z} \kern-.4em \hbox{\tenss Z} }

\def\L{ {\cal L} }

\def\ep{\epsilon}
\def\eps{\varepsilon^{\mu\nu\rho}}
\def\d{\partial}
\def\epdel{\,\varepsilon\partial\,}

\def\d{\partial}
\def\la{\raise.16ex\hbox{$\langle$} \, }
\def\ra{\, \raise.16ex\hbox{$\rangle$} }
\def\st{\, \raise.16ex\hbox{$|$} \, }
\def\go{\rightarrow}

\def\CS{ {\rm CS} }
\def\eff{ {\rm eff} }

\def\Dbar{ D \kern-.7em\raise.65em\hbox{$-$} }
\def\Dsquare{ D_k \kern-1.2em\raise.63em\hbox{$-{\scriptstyle 2}$} }
\def\Dbarky{ D_k \kern-1.2em\raise.63em\hbox{$-{\scriptstyle y}$} }
\def\DbarjI{ D_j \kern-1.2em\raise.63em\hbox{$-{\scriptstyle I}$} }
\def\DbarkI{ D_k \kern-1.2em\raise.63em\hbox{$-{\scriptstyle I}$} }

\def\psibar{ \psi \kern-.65em\raise.6em\hbox{$-$} }
\def\Lbar{ {\cal L} \kern-.65em\raise.6em\hbox{$-$} }

\def\ep{\epsilon}
\def\eps{\varepsilon^{\mu\nu\rho}}
\def\d{\partial}
\def\la{\raise.16ex\hbox{$\langle$} }
\def\ra{\raise.16ex\hbox{$\rangle$} }
\def\go{\rightarrow}

\def\psibar{ \psi \kern-.65em\raise.6em\hbox{$-$} }
\def\Dbar{ D \kern-.8em\raise.65em\hbox{$-$} }

\nopagenumbers
\rightline{December, 1995}
\rightline{Revised version}
\vskip 1.8cm

\baselineskip=22pt

\centerline{\bf Quantum Group Symmetry in Multiple Chern-Simons
Theory on a Torus}

\vskip 1.8cm

\baselineskip=13pt

\centerline{ Choon-Lin Ho}
\centerline{\sl Department of Physics, Tamkang University, Tamsui,
Taiwan 25137, R.O.C.}

\vskip 1.5cm

\baselineskip=16pt

\centerline{\bf Abstract}

\midinsert \narrower

We discuss $w_\infty$ and $sl_q(2)$ symmetries in multiple Chern-Simons
theory on a torus.
It is shown that these algebraic structures arise from the
dynamics of the non-integrable phases of the Chern-Simons fields.  The
generators of these algebras are constructed from the Wilson line operators
corresponding to these phases.  The vacuum states form the basis of cyclic
representation of $sl_q (2)$.

\endinsert

\vfill\eject

\baselineskip =24 truept
\pageno=2
\footline={\hss\tenrm\folio\hss}

In the last decade the studies of qauntum groups and algebras
have attracted a lot of interests. One can already see the great impact
these studies have in physics and mathematics.  The concept of
quantum groups and algebras has its origin in the development of the
quantum inverse method and the study of solutions to the Yang-Baxter
equation [1].  These new mathematical structures have already found
applications in
exactly solvable statistical models, in two-dimensional conformal field
theory and in non-abelian Chern-Simons (CS) theory [2].  Undoubtedly, it
is of interest to find applications
of these concepts in other important systems, especially realistic ones.

Recently a  $sl_q (2)$ quantum group symmetry is uncovered
in the Landau problem ({\it i.e.} charged
particle moving
in a constant magnetic field) and in the related problem of fractional
quantum Hall effects (FQHE) [3-8].
A quantum $w_\infty$ algebra, also known as the FFZ [9] algebra, is
also realised in these systems [4,5,10,11].
Representation of the quantum algebra $sl_q (2)$ was applied to formulate
the Bethe-{\sl ansatz}
for the problem of Bloch electron in magnetic field, {\it i.e.} the
Azbel-Hofstadter problem [3,12,13].

On the other hand, abelian Chern-Simons field theory with matter coupling
have attracted intense
interest in recent years, owing to its relevance to condensed matter
systems
such as  quantum Hall systems, and possibly  high $T_c$
superconductors.
Many studies have also been carried out for the Maxwell-Chern-Simons
theory in which a Maxwell kinetic term is included.
An interesting
observation is that the dynamics in the topological sector of
Maxwell-Chern-Simons theory on a
torus is equivalent to the Landau problem  on a torus.
Thus many interesting
features are shared by both systems.
In fact, the quantum algebras mentioned previously were
also found in the Maxwell and the pure CS theory on a torus [5,7].

In this paper we would like to extend these results
to a theory on a torus in which
multiple kinds of  Chern-Simons gauge interactions are introduced among
particles [14].
It has been known that multiple Chern-Simons interactions induce
matrix statistics which generalizes ordinary fractional statistics in
the space of particle species.
A possible application of the theory is in the double-layered Hall systems
[14,15].

Let us consider a theory on a torus (of lengths $L_1$ and $L_2$) with  $M$
distinct
CS gauge fields, $a_\mu^I$ ($I=1,\cdots,M$) and nonrelativistic
matter fields.
 The theory is described by the Lagrangian $\L = \L_\CS + \L_{\rm matter}$,
 where the Chern-Simons term $\L_\CS$ is given by
$$\eqalign{
 \L_\CS &= {1\over4\pi} \sum_{IJ} K_{IJ} ~ a^I \epdel a^J
   \hskip .6cm (I,J=1, \cdots, M)~~,  \cr
 &(\epdel a)^\mu \equiv \eps \d_\nu a_\rho ~~~. \cr}
   \eqno(1) $$
Here $K_ {IJ}$ is a $M\times M$ real symmetric matrix.
Properties of this theory is studied in [16], and we refer the reader
to this
reference for details.  Only those features relevant to the present
discussions are summarized below.

On the torus , for each CS field there are two nonintegrable phases,
$\theta^I_j$ ($j=1,2$), of the
Wilson line integrals along the two non-contractible loops of the torus.
These phases are new degrees of freedom undetermined by the matter
content, and their
contributions are found to be
decoupled in the action of the theory. They are responsible for the
 topological structures of the theory.  We will be interested only
in the structures of the
Hilbert space of these  Wilson line phases in the rest of the paper.

In [16] it is found that the Lagrangian corresponding to the Wilson
line
phases is in the form : $ {1\over 4\pi}\,  K_{IJ} ( \theta_2^I
\dot\theta_1^J -
      \theta_1^I \dot\theta_2^J ) $ .
This implies that $\theta^I_1$'s and $\theta^I_2$'s are conjugate pairs:
 $$[ \, \theta^I_1 , \theta^J_2 \,] =2\pi i \, K^{-1}_{IJ} ~~~,
       \eqno(2) $$
where $K^{-1}_{IJ}$ is the $IJ$-component of the matrix $K^{-1}$.
The system is invariant under
large gauge transformations which shift the Wilson line phases by
multiples of $2\pi$:
$$U^I_j ~:~~ \theta^I_j \go \theta^I_j + 2\pi ~~~.  \eqno(3) $$
Unitary operators inducing the transformation (3)
are given by
$$U^I_j  =  e^{+i \ep^{jk}\, K_{IJ} \theta^J_k}    ~~~. \eqno(4)$$

The two sets of operators, $\{ U^I_j \}$ and $\{ W^I_j \}$, are
complimentary.
They satisfy the Weyl-Heisenberg (WH) relations:
$$\eqalign{
\noalign{\kern 6pt}
U^I_1 ~ U^J_2 \, &= \, e^{-2\pi i K_{IJ}} ~ U^J_2\, U^I_1 ~~~, \cr
W^I_1 \, W^J_2 &= e^{-2\pi i K^{-1}_{IJ} } \, W^J_2\, W^I_1 ~~~, \cr
U^I_j ~ W^J_k &= W^J_k ~ U^I_j ~~~. \cr
\noalign{\kern 6pt}
}  \eqno(5)  $$
Note that these operators do not commute with each other in general.
The algebra is invariant under the interchange of $U^I_j$ and $W^I_j$
supplemented by the replacement of $K_{IJ}$ by $K^{-1}_{IJ}$.  This
suggests that there is a duality between the theories with the
 Chern-Simons coefficient matrix $K$ and with $K^{-1}$.

We would like to determine vacuum wave functions that form a
representation of the WH group (5).
From now on we shall suppose that all $K_{IJ}$'s are integers so that
all the $U_j^I$'s commute among themselves.     We may thus
simultaneously diagonalize these operators and take
$$U^I_j \, | \Psi \ra = e^{i \gamma^I_j} \, |\Psi\ra ~~~,
           \eqno(6)  $$
where $\gamma^I_j$ are the vacuum angles.
For convenience we introduce vector notation :
$\vtheta_j=(\theta^1_j, \cdots, \theta^M_j)$,
$\vgamma_j=(\gamma^1_j, \cdots, \gamma^M_j)$, etc.
It has been shown that the degeneracy of vacua is $r={\rm det} K$ [16,17].
An independent basis of vacua $| \vh_a \ra$ can be chosen as (in the
$\vtheta_1$ representation) [16]:
$$\eqalign{
\la \vtheta_1 | \,\vh_a \ra \equiv u_a(\vtheta_1)
  = e^{i \vgamma_1 \cdot \vtheta_1/2\pi}\,
  \delta_{2\pi} [\vtheta_1 + K^{-1}\, \vgamma_2 - \vh_a] ~~~,&\cr
(a=1, \cdots, ~ r= {\rm det} \,K) ~~~.& \cr}  \eqno(7) $$
The set of vectors ${\cal H}(K) = \{  \vh_a \}$ is defined by
$${\cal H} (K) = \{ \, \vh_a \in R^M , ~ (a=1, \cdots, r)
  ~;~ K\, \vh_a \sim 0 \, \} ~~, \eqno(8) $$
where the equivalence relation ~$\sim$~ among vectors $ \in R^M$
is defined by:
 $$\vh \sim \vg \quad \Longleftrightarrow \quad
 h^I = g^I \quad (mod~ 2\pi) \quad I=1, \cdots, M ~. \eqno(9) $$
Vectors in ${\cal H} (K)$  are independent in the
in the sense that $\vh_a \not\sim \vh_b$ iff $a\not= b$,

The actions of the Wilson lines on the vacuum are :
$$\eqalign{
W_1^I ~ \st \, \vh_a \ra
   &= e^{-i \vl_I \cdot \vgamma_2 - ih_a^I } ~ \st \, \vh_a \ra ~~, \cr
W_2^I ~ \st \, \vh_a \ra
   &= ~e^{+i \vl_I \cdot \vgamma_1} ~~ ~\st \, \vh_a -2\pi\vl_I \ra ~~.
   \cr}    \eqno(10)$$
Here $\vl_I$ are the column vectors of the matrix $K^{-1}$:
$K^{-1}  = ( \vl_1 \, , \cdots ,  \, \vl_M )$ .  Note that , if $\vk_I$
 are the column vectors of $K$ : $K ~= ( \vk_1 , \cdots , \vk_M )$ , then
 we have the
orthogonality relation:  $\vk_I \cdot \vl_J = \delta_{IJ} $.
It is easy to see that $W_2^I$ induces a mapping among the vacua.
In fact, we have
$K\, (\vh_a - 2\pi \, \vl_I) \sim 0$, so that
$\vh_a - 2\pi \vl_I \in {\cal H}$.  If there exists a $K^{-1}_{IJ}$ such
that $\exp(2\pi i K_{IJ}^{-1} )$ is a primitive $r^{th}$ root of unity ,
{\it i.e.} $r$ is the least integer such that $q^r =1$, then $(W_2^J)^r
\sim I$, and $W_2^J$ maps all the $r$ distinct vacua among themselves.

We can now reveal the qauntum algebraic structures inherent in the theory.
First let us form from the $W_j^I$ the following operators:
$$T_\n =T_{(n_1,n_2)}\equiv q^{n_1n_2/2} (W_1^I)^{n_1}
(W_2^J)^{n_2}~,\eqno(11)$$
where $q\equiv e^{2\pi i K_{IJ}^{-1} }$, and $n_1, n_2$ are integers.
Note here that there are $M\times M$ possible sets of operators $T$,
depending on the choice of $I$ and $J$.
From the WH algebras (5) one gets:
$$T_\m~T_\n = q^{-\m\times \n/2} T_{\m+\n}~,\eqno(12)$$
where $\m\times \n=m_1n_2-m_2n_1$.
Eq.(12) implies
$$\left[T_\m,T_\n\right]  = -2i
\sin\left({i\pi K_{IJ}^{-1} }\left(\m\times\n\right)\right)T_{\m+\n}~.
\eqno(13)$$
This is nothing but the quantum $w_\infty$ (FFZ) algebra.

Next we construct the operators $J_\pm$ and $J_3$ from the $T$'s as
follows [4-7]:
$$\eqalign{
J_+ &\equiv {1\over q-q^{-1}}\left(T_{(1,1)} - T_{(-1,1)}\right)~,\cr
J_- &\equiv {1\over q-q^{-1}}\left(T_{(-1,-1)} - T_{(1,-1)}\right)~,\cr
q^{2J_3} &\equiv T_{(-2,0)}~, \qquad\qquad
q^{-2J_3} \equiv  T_{(2,0)}~.\cr}\eqno(14)$$
Using (12), one can show that:
$$\eqalign{
q^{J_3}J_\pm q^{-J_3} &=q^{\pm 1}J_\pm~,\cr
[J_+,J_-]&=[2J_3]_q~,\cr}\eqno(15)$$
where $[x]\equiv (q^x-q^{-x})/(q-q^{-1})$.
This is the defining relations of the quantum algebra $sl_q (2)$, and
the $J$'s constructed according to (14) are just the generators of this
algebra.

We shall now show that the vacua form the basis of the representation
of the $sl_q (2)$ algebra.  To do this, we need to know the
actions of $T_\n$ on the state $\st \vh_a \ra $.
From the definition of $T_\n$ (11) and the WH algebra (5), we find :
$$
T_\n \st \vh_a \ra= q^{-{n_1n_2\over 2}}
e^{i\left(n_2\vl_J\cdot \vgamma_1 - n_1\vl_I\cdot \vgamma_2+ n_1 h_a^I
\right)} \st \vh_a - 2\pi n_2 \vl_J \ra~.
\eqno(16)$$
In obtaining (16), use has been made of the fact that the $I^{th}$
component of $\vl_J$ is $K_{IJ}^{-1}$.

With the help of (16), it is easy to obtain the actions of the $J$'s
on the state vectors $\st \vh_a \ra$ :
$$\eqalign{
J_+ \st \vh_a \ra &= e^{i\vl_J \cdot \vgamma_1}
\left[\left(2\pi K_{IJ}^{-1}\right)^{-1} \left(h_a^I - \vl_I
\cdot \vgamma_2\right)-{1\over 2}\right]_q  \st \vh_a - 2\pi \vl_J \ra~,\cr
J_- \st \vh_a \ra &= - e^{-i\vl_J \cdot \vgamma_1}
\left[\left(2\pi K_{IJ}^{-1}\right)^{-1} \left(h_a^I - \vl_I
\cdot \vgamma_2\right)+{1\over 2}\right]_q  \st \vh_a + 2\pi \vl_J \ra~,\cr
q^{\pm 2J_3} \st \vh_a \ra &=
     q^{\mp\left(\pi K_{IJ}^{-1}\right)^{-1}\left(h_a^I-\vl_I
\cdot \vgamma_2\right)} \st \vh_a \ra~,\cr}\eqno(17)$$

Since $W_2^J$ induces a mapping among the vacua as mentioned before,
the vacua $\vh_a$ will in general form a cyclic representation of
$sl_q (2)$ [18].
Suppose $K_{IJ}^{-1}$ is such that $q\equiv e^{2\pi i K_{IJ}^{-1} } $
is a $r^{th}$ root of unity, then the cyclic representation is
irreducible, and is of demension $r$.   Highest weight representation,
 however, could sometimes be obtained with an appropriate choice of
the vacuum angles, as discussed in [7].

We now apply these results to two cases relevant to the FQHE.

\vskip 1cm
\leftline{\tt Case I. ~ $K=\left(\matrix{ 3&2\cr 2&3\cr} \right)$}

\vskip 8pt
A theory defined by this $K$ matrix serves as an alternative way of
describing the first daughter state in the FQHE.
This $K$ gives
$$K^{-1}={1\over 5} \left(\matrix{ 3&-2\cr -2&3\cr} \right) ~~~,~~~
  {\rm det}\, K=5 ~~~.
   \eqno(18) $$
The  $ \vh_a $ for the vacua in (7) is chosen to be
$$\vh_a = {2\pi a\over 5} ~ \vone \hskip .6cm (a=0, \cdots, 4) ~.
    \eqno(19)  $$

For simplicity, we label the vacua $\st \vh_a\ra$ by $\st a \ra$.
They satisfy
$\st a+5 \ra = \st a \ra ~$.
Vectors $\vl_I$'s are given by
$$\vl_1 = {1\over 5} \left( { 3\atop -2} \right) \next
  \vl_2 = {1\over 5} \left( { -2\atop 3} \right) ~~~,\eqno(20) $$
and
$$\vh_a - 2\pi \vl_I \sim \vh_{a+2} ~~~,   \eqno(21)$$
From (10), we have,
$$\eqalign{
W_1^I ~ \st  a \ra
 &= e^{-i \vl_I \cdot \vgamma_2 + 2\pi ia/5} ~ \st  a \ra ~~, \cr
W_2^I ~ \st  a \ra
 &= ~~ e^{+i \vl_I \cdot \vgamma_1} ~~~ \st  a+2 \ra ~~. \cr}
    \eqno(22) $$

As $M=2$ in this case, there are altogether four possible ways of
forming the $T$ operators according to (11), and therefore there
 exist four possible
sets each of the quantum $w_\infty$ and $sl_q (2)$ algebras.
Note here that
all the $q$'s formed from the four $K^{-1}_{IJ}$ are primitive
 $r^{th}$ root of unity.
Suppose we form the operators $T_\n$ with $I=1$ and $J=2$.  Then
$K^{-1}_{12}=-{2\over 5}$ and $q=e^{-4\pi i /5}$.  The actions of
$J_{\pm,3}$ as defined in (14) on the vacuum states are evaluated
to be:
$$\eqalign{
J_+ \st a \ra &= e^{i{\beta}_1/5}
\left[-{a\over 2}+{\alpha_2\over 4\pi}-{1\over
2}\right]_q  \st a+2 \ra~,\cr
J_- \st a \ra &=- e^{-i{\beta}_1/5}
\left[-{a\over 2}+{\alpha_2\over 4\pi}+{1\over
2}\right]_q  \st a-2 \ra~,\cr
q^{\pm 2J_3} \st a \ra &=
     q^{\pm 2 \left({a\over 2}-{\alpha_2\over 4\pi}\right)}
\st a \ra~,\cr}\eqno(23)$$
where $\alpha_i\equiv 3\gamma^{(1)}_i - 2\gamma^{(2)}_i$ and
$\beta_i\equiv -2\gamma^{(1)}_i + 3\gamma^{(2)}_i$

It turns out that in this case there is, in addition to the four
possible sets
of $J$'s mentioned before, another set of generators of $sl_q (2)$
 algebra. This is seen
as follows.  In [16] it is shown that, under appropriate coupling
with matter
fields, the multiple CS theory is effectively  equivalent to a single
CS theory with effective Chern-Simons coefficient given by
$\N_\eff^{-1} = \sum_{I,J} K^{-1}_{IJ}~ $.
$\kappa_{\rm eff}$ is in general a rational number, $\N_\eff =p/q$
( $p$ and $q$ are two mutually prime integers), even if all $K_{IJ}$'s are
integers.  The effective Wilson line operators $\bW_i$ are related to the
$W^I_i$ in the multiple CS theory by
$\bW_i = W^{(1)}_i W^{(2)}_i$, and satisfy the WH algebra:
$$\bW_1 \bW_2 = \bq^{-1}  \bW_2 \bW_1 ~~~ ,
\qquad\qquad\qquad \bq\equiv e^{2\pi i /\N_\eff}.\eqno(24)  $$
In the present case, one has $\N_\eff=5/2$.  The actions of $\bW$'s on
$\st a \ra$ are:
$$\eqalign{
\bW_1 \, \st a \ra &= e^{- i(\lambda_2 - 4\pi a)/5 } \, \st a \ra ~~~,\cr
\bW_2 \, \st a \ra &= ~e^{i\lambda_1 /5 } \, ~\st a-1 \ra ~~~, \cr}
  \eqno(25) $$
where $\lambda_j\equiv \gamma^{(1)}_j + \gamma^{(2)}_j$.
One can now define a new set of $sl_\bq (2)$ generators $\bJ$'s
 according to (14),
with the $W$ 's replaced by the $\bW$'s.  The actions of $\bJ$ is
found to be:
$$\eqalign{
\bJ_+ \st a \ra &= e^{i\lambda_1/5}
\left[a-{\lambda_2\over 4\pi}-{1\over 2}\right]_\bq  \st a-1 \ra~,\cr
\bJ_- \st a \ra &=- e^{-i\lambda_1/5}
\left[a-{\lambda_2\over 4\pi}+{1\over 2}\right]_\bq  \st a+1 \ra~,\cr
\bq^{\pm 2J_3} \st a \ra &=
    \bq^{\mp 2 \left(a-{\lambda_2\over 4\pi}\right)}
\st a \ra~,\cr}\eqno(26)$$
These expressions are precisely those obtained in [5,7]
for the case of single CS theory.

\vskip 1cm
\vskip 8pt
\leftline{\tt Case II. ~ $K= \left(\matrix { 5&3\cr 3&3\cr} \right)$}

\vskip 8pt

This case corresponds to a filling factor
$1/3$ in the FQHE.
We have in this case
$$K^{-1} = {1\over 6}\left(\matrix{ 3&-3\cr -3&5\cr} \right) ~~~,~~~
{\rm det}\, K = 6 ~~~,~~~ \N_{\rm eff}={3} ~~~.  \eqno(27) $$
A choice $\{ \vh_a \}$ for the vacua can be chosen to be
$$\vh_a =  {\pi a\over 3} ~ \left( {3\atop -5}
  \right) \qquad (a=0 \sim 5)~.   \eqno(28)$$
We note that
$\st a+6 \ra = \st a \ra~$.
This time we have
$$\vl_1 = {1\over 2} \left( { 1\atop -1} \right) \next
  \vl_2 = {1\over 6} \left( { -3\atop 5} \right)
     \eqno(29) $$
so that
$$\eqalign{
&\vh_a - 2\pi \vl_1 \sim \vh_{a-3} ~~~,\cr
&\vh_a - 2\pi \vl_2 \sim \vh_{a+1} ~~~. \cr}   \eqno(30) $$
Hence the actions of Wilson line operators are given by (we set the
vacuum angles $\vgamma_i =0 $ for simplicity):
$$\eqalign{
W_1^{(1)} ~ \st  a \ra
 &= e^{i\pi a} ~ \st  a \ra ~~, \cr
W_1^{(2)} ~ \st  a \ra
 &= e^{-5i\pi a/3} ~ \st  a \ra ~~, \cr
W_2^{(1)} ~ \st  a \ra
 &= ~~\st  a-3 \ra ~~. \cr
W_2^{(2)} ~ \st  a \ra
 &= ~~\st  a+1 \ra ~~. \cr}
    \eqno(31) $$

As discussed previously, since only $q\equiv e^{2\pi i K_{IJ}^{-1} } $
with $K_{IJ} = K_{22}$ is a primitive $r^{th}$ root of unity, the
six states only
form the basis of an irreducible cyclic representation of a $sl_q (2)$
algebra for the $J$'s constructed by $W_1^{(2)}$  $W_2^{(2)}$.
The actions
of $J_{\pm,3}$ so constructed on the vacuum states are evaluated to be:
$$\eqalign{
J_+ \st a \ra &=
-\left[a+{1\over 2}\right]_q  \st a+1 \ra~,\cr
J_- \st a \ra &=
\left[a- {1\over 2}\right]_q  \st a-1 \ra~,\cr
q^{\pm 2J_3} \st a \ra &=
     q^{\pm 2 a}~\st a \ra~,\cr}\eqno(32)$$

We note here that, unlike case (I), the vacua $\st a \ra$ in this case
do not
form the basis of irreducible cyclic representation for operators $\bJ$'s
obtained
from the $\bW_i= W^{(1)}_i W^{(2)}_i$ .  It is easily checked that
$\bq=e^{2\pi i/3}$ ($\bq^3=1$).  Thus there are only three
inequivalent
states in the irreducible cyclic representation of the $sl_\bq$ algebra
generated by the $\bJ$'s.
This is related to the fact, as discussed in [16], that there exist only
three inequivalent states in the effective single CS theory with
$\N_\eff =3$.
The actions of $\bW_2$ on $\st a \ra$ is :
$\bW_2 \st a \ra = \st a-2 \ra$, which seperate the six states
into two groups:
$\{ \st a\ra ; a=0,2,4 \}$ and $\{ \st a\ra ; a=1,3,5 \}$.  On the
other hand,
by a similar argument given in [16], one can check that the states
$\st a\ra$ and
$\st a+3 \ra$ are equivalent in the effective theory.  That means the
states
$\{\st 0\ra, \st 3\ra \}$,  $\{\st 1\ra, \st  4\ra \}$ and
$\{\st 2\ra, \st 5\ra \}$
correspond to the three distinct states in the effective theory.
The Wilson line operator $\bW_2$, and hence the $\bJ_\pm$'s, map
 states among these three groups.

Finally, we remark that the above discussions and results can be directly
carried over to the Maxwell-Chern-Simons theory with multiple kinds of CS
fields and the related fractional quantum Hall theory on a
torus [17].  When
the Maxwell terms are included, the relevant operators are the
so-called
magnetic translation operators.  Ground states form the basis of an
algebra satisfied by these translation operators, which are
 precisely the WH group (5)
obeyed by the $W^I_j$'s.  Thus, while the actual forms of the
ground states
differ in the two theories, the algebraic structures are exactly
the same. So they also share the same quantum group structures
discussed in this paper.

\vskip 1. truecm

\centerline{\bf Acknowledgement}

This work is supported by R.O.C. Grant NSC 85-2112-M-032-002.

\vfil\eject

\def\ijmpA#1#2#3{{\it Int.\ J.\ Mod.\ Phys.} {\bf {A#1}}, #3 (19{#2})}
\def\ijmpB#1#2#3{{\it Int.\ J.\ Mod.\ Phys.} {\bf {B#1}}, #3 (19{#2})}
\def\jmp#1#2#3{{\it  J.\ Math.\ Phys.} {\bf {#1}}, #3 (19{#2})}
\def\mplA#1#2#3{{\it Mod.\ Phys.\ Lett.} {\bf A{#1}}, #3 (19{#2})}

\def\plB#1#2#3{{\it Phys.\ Lett.} {\bf {#1}B}, #3 (19{#2})}

\def\np#1#2#3{{\it Nucl.\ Phys.} {\bf B{#1}}, #3 (19{#2})}
\def\prl#1#2#3{{\it Phys.\ Rev.\ Lett.} {\bf #1}, #3 (19{#2})}

\def\jpA#1#2#3{{\it J.\ Phys.} {\bf A{#1}}, #3 (19{#2})}

\def\ptp#1#2#3{{\it Prog.\ Theor.\ Phys.} {\bf {#1}}, #3 (19{#2})}
\def\smd#1#2#3{{\it Soviet\ Math.\ Dokl.} {\bf {#1}}, #3 (19{#2})}
\def\lmp#1#2#3{{\it Lett.\ Math.\ Phys.} {\bf {#1}}, #3 (19{#2})}
\def\jsm#1#2#3{{\it J.\ Sov.\ Math.} {\bf {#1}}, #3 (19{#2})}
\def\tmp#1#2#3{{\it Theor.\ Math.\ Phys.} {\bf {#1}}, #3 (19{#2})}

\parindent=15pt

\centerline{\bf References}

\item{[1]}
V. Drinfeld, \smd {32} {85} {254};
M. Jimbo, \lmp {10} {85} {63};
P. Kulish and N. Reshetikhin, \jsm {23} {83} {2435};
E. Sklyanin, L. Takhatajan and L. Faddeev, \tmp {40} {79} {688}.

\item{[2]}
H.J. de Vega, \ijmpA {4} {89} {2371};
L. Alvarez-Gaume, C. Gomez and G. Sierra, \np {319} {89} {155};
G. Siopsis, \mplA {6} {91} {1515}.

\item{[3]}
P.B. Wiegmann and A.V. Zabrodin, \prl {72} {94} {1890};
\np {422} {95} {495};
cond-mat/9310017.

\item{[4]}
H.-T. Sato, \mplA {9} {94} {451}; {\it ibid.} 1819; \ptp  {93} {94} {195}.

\item{[5]}
I.I. Kogan, \mplA {7} {92} {3717};  \ijmpA {9} {94} {3889}.

\item{[6]}
G.-H. Chen, L.-M. Kuang and M.-L. Ge, cond-mat/9509091;
G.-H. Chen and M.-L. Ge, cond-mat/9509092.

\item{[7]}
C.-L. Ho, {\sl $w_\infty$ and $sl_q (2)$ algebras in the Landau
problem and Chern-Simons theory on a torus}, Tamkang prprint, 1995.

\item{[8]}
M. Alimohammadi and A. Shafei deh Abad, hep-th/9503117.

\item{[9]}
D.B. Fairlie, P. Fletcher and C.K. Zachos, \plB {218} {89} {203};
\jmp {31} {90} {1088};
E.G. Floratos, \plB {228} {89} {335}; {\bf 232B} 467 (1989).

\item{[10]}
T. Dereli and A. Vercin, \plB {288} {92} {109}; \jpA {26} {93} {6961}.

\item{[11]}
S. Iso, D. Karabali and B. Sakita, \plB {296} {92} {143};
B. Sakita, \plB {315} {93} {124};
A. Cappelli, C.A. Trugenberger and G.R. Zembra, \np {396} {93} {465};
A. Cappelli, G.V. Dunne, C.A. Trugenberger and G.R. Zembra, \np {398} {93}
{531};
J. Martinez and M. Stone, \ijmpB {7} {93} {4389};
H. Azuma, \ptp {92} {94} {293}.

\item{[12]}
L.D. Faddeev and R.M. Kashaev, hep-th/9312133.

\item{[13]}
Y. Hatsugai, M. Kohmota and Y.S. Wu, \prl {73} {94} {1134};
cond-mat/9509062.

\item{[14]}
For a recent review, see eg.:A. Zee, {\sl Quantum Hall Fluids},
UCSB preprint NSF-ITP-95-08, 1995.

\item{[15]}
M. Greiter, X.-G. Wen and F. Wilczek, \prl {66} {91} {3205};
X.-G. Wen and A. Zee, \prl {69} {92} {1811}.

\item{[16]}
D. Wesolowski, Y. Hosotani and C.-L. Ho, \ijmpA {9} {94} {969}.

\item{[17]}
E. Keski-Vakkuri and X.-G. Wen, \ijmpA {7} {93} {4227}.

\item{[18]}
V. Pasquier and H. Saleur, \np {330} {90} {523};
E.G. Floratos, \plB {233} {89} {395};
Z.-Q. Ma, {\sl Yang-Baxter Equation and Quantum Enveloping Algebras},
 World Scientific, Singapore, 1993;
L.G. Biedenharn and M.A. Lohe, {\sl Quantum Group Symmetry and
$q$-tensor  Algebras},   World Scientific, Singapore, 1995.

\vfill
\bye